\documentclass[12pt]{article} 
\usepackage{amssymb}
\usepackage{graphicx}
\begin{document} 
\begin{center}

{\bf Thresholds of ultraperipheral processes}

\vspace{2mm}

I.M. Dremin\footnote{e-mail: dremin@lpi.ru}\\

{\it Lebedev Physical Institute, Moscow, Russia}

\end{center}

                                             
\vspace{1mm}

\begin{abstract}
Threshold behavior of the cross sections of ultraperipheral nuclear
interactions is studied. Production of $e^+e^-$ and $\mu ^+\mu ^-$ pairs
as well as $\pi ^0$ and parapositronium is treated. The values of
corresponding energy thresholds are presented and the total cross sections
of these processes at the newly constructed NICA and FAIR facilities are
estimated.
\end{abstract}

\vspace{1mm}


\vspace{1mm}

The high energy heavy ion colliders RHIC and LHC provide important information
on properties of strong interactions in head-on ion collisions. The problem
of quark-gluon plasma is widely discussed. Besides, the ions traversing at
large transverse distances (impact parameters) from one another interact by
their electromagnetic fields extended well beyond the range of nuclear forces.
Namely these processes are called ultraperipheral ones.

Production of electron-positron pairs by these fields was studied by
Landau and Lifshitz \cite{lali} as early as in 1934. The peculiar 
asymptotical energy dependence of the cross section of this process
was demonstrated. It increases with increasing energy as $\ln ^3\gamma $
at $\gamma \gg 1$ where $\gamma $ is the Lorentz-factor of colliding ions
$\gamma =E/M$ ($E$ and $M$ are their energies and masses). This dependence
overshoots the limit for asymptotical behavior of the total cross section of
strong (QCD) interactions which is $\ln ^2\gamma $ according to the famous
Froissart theorem \cite{froi}. The electromagnetic fields of heavy ions
are so strong that the effective coupling factor appearing in ultraperipheral 
cross sections $Z^4\alpha ^4\approx 0.13$ for Pb-ions with $Z=82$ becomes 
comparable to the strength of QCD forces. Rapid increase of these
cross sections with increasing energy makes them compatible with hadron cross 
sections \cite{uoh}, at least in some parts of the phase space volume due to 
specific kinematics of ultraperipheral processes.

Thus, ultraperipheral production of lepton-antilepton pairs becomes noticeable
at high energies and is experimentally studied already at RHIC and LHC 
\cite{ada, chat, abb, kha, at1, at2, at3}. Experimental results are
extensively compared with theoretical predictions (see review papers
\cite{bgms, beba, baur, bekn, baht, rvx, ufnd}). It is shown
\cite{vyzh, agm, szcz, dys} that experimental fiducial\footnote{
Measured within the limited phase space defined by experimental conditions.} 
cross sections agree quite well with theoretical predictions obtained  in the
framework of the equivalent photon approximatiom.

The designed ion energies at the newly constructed facilities NICA in Russia
(up to $\sqrt {s_{nn}}$=9 GeV per a nucleon pair in AuAu-mode and 12.5 GeV 
in pp-mode) and FAIR in Germany (7 GeV) are much lower 
than at RHIC and LHC ($10^2 - 10^4$ GeV). Therefore it is important to estimate
the energy thresholds of cross sections of various ultraperipheral processes
and choose those admissible for experimentation at these colliders. This is
the main goal of the present brief contribution.

Feynman diagrams of ultraperipheral processes contain the two-photon 
interactions leading to production of some final states (e.g., $e^+e^-$ pairs).
This blob can be described by the cross sections of these processes.
The missing element of the whole picture are the photon fluxes emitted by
the colliding charged objects. This problem was first solved by Fermi
\cite{fer1, fer2} in 1924 and used by Weizs\"{a}cker \cite{wei} and Williams
\cite{wil} in 1934 who formulated the equivalent photon approximation.
The photon fluxes are calculated from the flux of electromagnetic fields
surrounding colliding ions. They are shown in many textbooks on quantum 
electrodynamics. 

The photons carrying small fractions $x$ of the nucleon energy
dominate in these fluxes. The~distribution of equivalent photons generated by a moving
nucleus with the charge $Ze$ and carrying a fraction of 
the nucleon energy $x$ integrated over the transverse momentum up to some value 
(see, e.g., \cite{blp}) looks as
\begin{equation}
\frac {dn}{dx}=\frac {2Z^2\alpha }{\pi x}\ln \frac {u(Z)}{x}.\label{flux}
\end{equation}
The ultraperipherality parameter $u(Z)$ depends on the nature of colliding 
objects and created states. Its physical meaning is the ratio of the maximum 
adoptable transverse momentum to the nucleon mass as the only massless 
parameter of the problem. It differs numerically in various approaches 
\cite{bgms, vyzh, bb1, kn, bkn, balt, kgsz, seng, zha, kmr, geom}. 
It depends on charges $Z_ie$, sizes and impact parameters of colliding objects
(form factors and absorptive factors) as well as, in principle, on considered 
processes. 

The essence of the equivalent photon approximation consists in convolution
of these spectra with total cross sections of corresponding processes 
$\gamma \gamma \rightarrow X$. Then
the exclusive cross section of production of the final state $X$ 
in the two-photon collisions of nuclei $A$ is written as
\begin{equation}
\sigma _{AA}(\gamma \gamma \rightarrow X)=
\int dx_1dx_2\frac {dn}{dx_1}\frac {dn}{dx_2}\sigma _{\gamma \gamma }(X),
\label{e2}
\end{equation}
where the fluxes $dn/dx_i$ for the colliding objects 1 and 2
are given by Eq. (\ref{flux}).

Creation of $\mu ^+\mu ^-$ pairs is rather well studied at LHC both in pp-
and PbPb-collisions. Experimental
data are confronted with theoretical results in the equivalent photon
approximation in papers \cite{vyzh, agm, szcz}. Using Eqs (\ref{flux}) and
(\ref{e2}) one gets
\begin{equation}
\sigma _{AA}(\gamma \gamma \rightarrow \mu ^+\mu ^-)=
\frac {2Z^4\alpha ^2}{3\pi ^2}\int _{4m^2_{\mu }}^{u^2s_{nn}}\frac {ds_0}{s_0} 
\sigma _{\gamma \gamma }(\mu ^+\mu ^-)\ln ^3\frac {u^2s_{nn}}{s_0}.
\label{mu2}
\end{equation}
The cross section $\sigma _{\gamma \gamma }(\mu ^+\mu ^-)$ to be inserted 
in (\ref{mu2}) looks \cite{brwh, blp} as
\begin{equation}
\sigma _{\gamma \gamma }(\mu ^+\mu ^-)=\frac {\pi \alpha ^2}{2m^2_{\mu }}
(1-z^2)[(3-z^4)\ln \frac {1+z}{1-z}-2z(2-z^2)], 
\label{mu}
\end{equation}
where $z=\sqrt {1-\frac {4m^2_{\mu}}{s_0}}$ and $s_0$ is the squared invariant
mass of two muons.

This cross section tends to 0 at the threshold $s_0=4m^2_{\mu}$ and decreases
as $\frac {1}{s_0}\ln s_0$ at very large $s_0$. Therefore, with logarithmic 
accuracy, some finite value can be inserted in the $\ln ^3$-term in (\ref{mu2}).
Its choice is unimportant for estimates at very high energies. However, for our 
purposes this value should be chosen close to the maximum of (\ref{mu})
near $s_0\vert _{eff}=8m^2_{\mu}$. Then the ultraperipheral cross section 
looks as
\begin{equation}
\sigma _{AA}(\gamma \gamma \rightarrow \mu ^+\mu ^-)=
\frac {28Z^4\alpha ^4}{27\pi m^2_{\mu }}\ln ^3\frac {u^2s_{nn}}{8m^2_{\mu }}.
\label{mu3}
\end{equation}
The typical $\ln ^3\gamma $ asymptotical dependence is clearly seen. At the 
same time, it follows that the unbound pairs of muons can be created only
at energies exceeding the threshold value
\begin{equation}
\sqrt {s_{nn,thr}}=\sqrt 8 m_{\mu}/u.
\label{thr} 
\end{equation}
This effective threshold can be evaluated if the value of the 
ultraperipherality parameter $u$ for heavy ions is known.
In the paper \cite{vyzh} it is estimated as $u_{PbPb}\approx 0.02$ for Pb-ions
(it is practically the same for Au-ions). 
One concludes that unbound muon pairs are produced by such mechanism only at 
energies (per a pair of nucleons) exceeding about 15 GeV. This 
threshold is higher than energies which will be available at NICA or FAIR.

Thus, the only hope left is the production of electron-positron pairs or
low-lying resonances. The resonance $R$ cross section is written as
\begin{equation}
\sigma _{\gamma \gamma }(R)=\frac {8\pi ^2\Gamma _{tot}(R)}{m _R}
Br(R\rightarrow \gamma \gamma )Br_d(R)\delta (x_1x_2s_{nn}-m_R^2).
\label{e3}
\end{equation}
Here, $m_R$ is the mass of $R$, $\Gamma _{tot}(R)$ its total width and  
$Br_d(R)$ denotes the branching ratio to a considered channel of its decay.
$s_{nn}=(2m\gamma )^2, m$ is a nucleon mass.
The $\delta $-function approximation is used for resonances with small
widths compared to their masses.

The total ultraperipheral cross section of resonance production looks 
\cite{geom} as
\begin{equation}
\sigma _{AA}(R)=\frac {128}{3}Z^4\alpha ^2Br(R\rightarrow \gamma \gamma )Br_d(R)
\frac {\Gamma _{tot}(R)}{m_R^3}\ln ^3\frac {u\sqrt {s_{nn}} }{m_R}.
\label{e4}
\end{equation}
The asymptotical $\ln ^3\gamma $ behavior is valid again. The factor $2um/m_R$ 
defines both threshold and preasymptotical behavior of the 
ultraperipheral cross section of resonance production.
As expected, the energy dependence in Eq. (\ref{mu3}) for (yet unobserved) 
dimuons with $m_R\approx 2m_{\mu }$ is the same as in (\ref{e4}) if
the effective value $s_0=4m^2_{\mu }$ was used in (\ref{mu2}). It is easy
to see that dimuons can not be produced at NICA and FAIR.

The only available channels are either the electron-positron pairs including
parapositronium or $\pi ^0$. For parapositronium with $m_{Ps}=1.022$ MeV
and $\Gamma _{Ps}=5.1\cdot 10^{-12}$ MeV the evaluated total cross sections
are about 22 mb at NICA and 19 mb at FAIR. For $\pi ^0$ with 
$m_{\pi ^0}=135$ MeV and $\Gamma _{\pi ^0}=7.5\cdot 10^{-6}$ MeV
they are much less
at NICA (about 2.3 $\mu $b) and very close to the threshold at FAIR.

The special kinematical features of these resonances produced with low
velocities can help in resolving them by registration of gamma-quanta
with discrete energies 0.511 MeV from the decay of parapositronia and
67.5 MeV from $\pi ^0$. Special detectors must be installed for these
searches. Then the corresponding fiducial cross section can be calculated.
The total cross sections for unbound electron-positron pairs are approximately 
the same as for parapositronia. However, their detection asks for other
detectors.
 
The ultraperipherality parameter $u$ is the most indefunite element of the 
whole approach. Threshold and preasymptotical behavior of cross sections
are defined by this parameter. It results from integration of the energy flux
over transverse momenta $q_t$ such that the formula (\ref{flux}) is determined 
by the form factors $F$ of colliding ions in a following way:
\begin{equation}
\frac {dn}{dx}=\frac {Z^2\alpha }{\pi ^2x}\int \frac {q^2_tF^2(q^2_t+4m^2x^2)}
{(q^2_t+4m^2x^2)^2}dq^2_t.
\label{flux1}
\end{equation}
The form factors cut transverse momenta and the integral converges.  The 
parameter $u$ is directly related with parameters of corresponding form factors.
Heavy ions are less stable than protons. Therefore smaller transverse momenta
are admissible for them. Correspondingly, their ulraperipherality parameter
$u_{PbPb}\approx 0.02$ is about ten times smaller than for protons 
$u_{pp}\approx 0.2$. These results follow from (\ref{flux1}) if the 
form factors of protons and nuclei are inserted there \cite{vyzh}.
The $q_t$-dependence of form factors is extracted from measurements of charge 
distributions in ions by Fourier transformation. The accuracy of the estimate 
of the ultraperipherality parameter depends on the precision of the data used
and on cuts of transverse momenta. The spatial pattern of the effective 
impact parameters for different processes imposes furher restrictions.
Therefore, unfortunately, one can rely on the above numerical values up to
the factor about 1.5. That shows the flexibility of the threshold estimates too.

The thresholds are inverse proportional (\ref{thr}) to the ultraperipherality 
parameter. They are lower for pp-modes. Then the $\ln ^3$-term becomes larger.
However this effect and higher energies about 12.5 GeV designed at NICA can 
not compensate the loss of the factor $Z^4$. The cross sections are much 
smaller in the pp-mode than estimated above for Au ions.

To conclude, it is a hard but not hopeless task to observe byproducts 
of ultraperipheral AuAu-collisions at NICA and FAIR energies. They can reveal 
themselves as monochromatic photons from decays of positronia and $\pi ^0$.
If observed, they would signal validity of theoretical approaches 
to their description even at rather low energies.

\vspace{6pt}

{\bf Acknowledgement}

This work was supported by the RFBR project 18-02-40131.

\end{document}